# Preliminary design of a NIR prime focus corrector for the Galileo Telescope

Claudio Pernechele[a], Favio Bortoletto[a,b], Paolo Conconi[c],
Daniele Gardiol[b], Emilio Molinari[c], Filippo Zerbi[c]

[a]Osservatorio Astronomico di Padova, vicolo dell'osservatorio 5, I-35100 Padova – Italy.

[b]Centro Galileo Galilei, Apartado de Correos 565, E-38700 S.ta Cruz de La Palma – Spain.

[c]Osservatorio Astronomico di Milano, via Bianchi 46, I-22055 Merate – Italy.

## ABSTRACT

In this paper a preliminary design for a prime focus corrector to be mounted at the Telescopio Nazionale Galileo (TNG) is presented. The telescope is located on La Palma (Canary Islands) and has a primary mirror of 3.5 m with a hyperbolic sag. Two optical designs have been considered in order to exploit detectors with 1k x 1k and 2k x 2k pixels. Each design makes use of four lenses, two of which are aspherical. The first lens diameters are, respectively, of 140 mm and 320 mm for the two kind of detectors. For both designs the telescope pupil is deliberately not re-imaged, and therefore it will not be possible to insert a cold stop. For such a reason particular care has been dedicated to the telescope thermal background study, in order to optimise the baffling system. The optics is able to correct fields of 11' x 11' and of 60'x60', depending on the design. It should be considered that the particular geometry of the focal plane array mosaic does not permit a full exploitation of the entire field, being based on a combination of four detectors. The selected chips are HgCdTe manufactured by Rockwell, i.e. Hawaii I in the 1k x 1k case and Hawaii II in the 2k x 2k case.

## 1. INTRODUCTION

In this report a preliminary study for a prime focus corrector for the Galileo Telescope in the near IR (NIR, 1.0 – 2.5 μm) spectral range is presented. In this study no scientific objective or detailed constructions aspects are taken into account. Two kind of optical layouts have been studied, both based on 2 x 2 detector mosaics. Being the detectors the elements with the mayor cost impact, the choice of the final optical design will be strongly driven by availability and cost of the focal plane array.

### 1.1 TNG primary mirror main parameters

The main optical parameters of the TNG primary mirror are listed in the following Tab. 1. The telescope has an aplanatic Cassegrain configuration with a hyperbolic primary mirror F/2.2 of 3.58 m diameter. The focal ratio of the primary mirror induces a naked scale on prime focus of 27 arcsec/mm. In order to prevent from high lens curvature, a slightly afocal corrector is here suggested, implying small variation on the naked scale. Considering a pixel size of about 18 μm (typical for a nowadays NIR detector), a "naked" scale of about 0.48 arcsec/px should then be expected. In the following pages we propose an optical design with a focal ratio of F/2.4 providing a slightly better sampling of 0.44 arcsec/pixel.

| Parameter | Value |
|---|---|
| Diameter | 3500 mm |
| Focal length | 7700 mm |
| Curvature radius | 15400 mm |
| Conic (hyperbolic) constant (K1) | -1.023818 |
| F/# | F/2.2 |
| Scale on prime focus | 27 arcsec/mm |

**Tab. 1:** TNG primary mirror main optical parameters.

## 1.2 The site of La Palma

The diffraction disk diameter of the TNG (3.58 m) is 0.07 arcsec at a wavelength of 500 nm, and 0.35 arcsec at the upper limit of the NIR range, i.e. 2.5 μm. The telescope is sited on La Palma Island, where measurements of seeing made by the IAC (Instituto de Astrofisica de Canarias) show a typical seeing disk diameter of about 0.6–1.0 arcsec all along the year, with a median value (measured during the year 1996) of 0.7 arcsec[1]. This result is confirmed by several observations at TNG, performed just after its first light (summer 1998). Such a seeing disk at 500 nm corresponds to a disk of about 0.5 arcsec at 2.5 μm.

Several authors have tried to estimate the goodness of sampling in astronomical observations. In the case of stellar photometry it has been suggested the use of the merit parameter R=FWHM/px, where FWHM is the full width at half maximum of the seeing disk (we fix it to 0.8 arcsec) and px the pixel dimension measured in the same unit. Some authors[2] have shown that the signal to noise ratio (SNR) of an acquisition as function of the parameter R has a maximum around 1. Following these authors, a sampling between 0.4 and 0.8 arcsec would be optimal, keeping in mind that the sampling should be a compromise with a field as wide as possible. However, it is generally recognised that the seeing disk has to be sampled at the Nyquist frequency: in our case the optimum pixel coverage is 0.35 arcsec/px.

On the other hand we have chosen to maintain the optical design as simple as possible and a better sampling implies greater lenses curvatures and higher aspherical coefficients. For this reason the final compromise for the design described in this report will be of 0.44 arcsec/px.

## 2. OPTICAL DESIGNS

Based on the considerations described in the precedent paragraph, we present here two different kinds of optical designs, both based on a typical detector with pixel size of 18 μm and a final focal ratio of F/2.2 and F/2.4 respectively.

### 2.1 Mosaic geometry.

Presently commercial NIR detectors are not "abuttable" like several more conventional CCD chips, so a considerable detector inter-space has to be considered in the design. For such a reason we have chosen to build the mosaic optimising the number of acquisitions needed to reach a given field of view. As an example, in Fig. 1 a 4 × 2k × 2k mosaic option is shown. In this case the detector inter-space is of 30 mm allowing to cover, with a slight superposition, 1 square degree of sky in four acquisitions. Each exposure has to be sinchronyzed with a correspondent telescope slew in the field.
The superposition simplifies the process of composite image calibration and reconstruction.

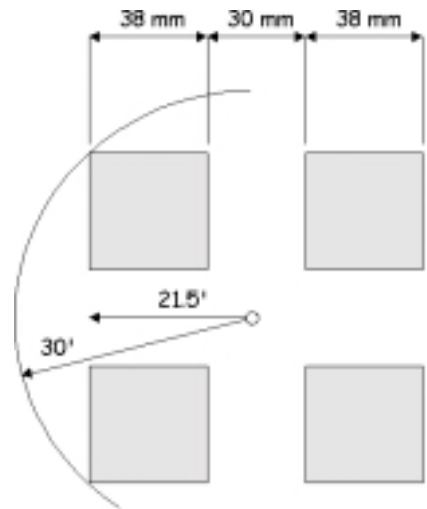

**Fig. 1:** A possible design for 4 chips 2k x 2k mosaic.

### 2.2 Option with 4 x 1k x 1k mosaic.

In the option 1 a mosaic of 4 × 1k × 1k chips and a completely afocal camera (f=7700 mm, F/#=2.2) has been chosen. The field of view is of about 24' × 24'. The 80% of the encircled energy is contained in 10 μm, well down the limit of the pixel sample (18 μm). The distance between the last lens (coincident with the camera window) and the mosaic is of 24 mm, but a larger distance is also possible. The diameter of the first lens is of 136 mm, while the length of the whole instrument is about 200 mm.

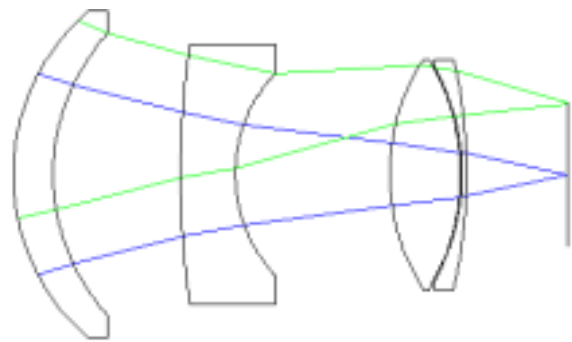

**Fig 2:** Optical layout for the 4 x 2k x 2k mosaic option.

## 2.3 Option with 4 x 2k x 2k mosaic.

This option allows the maximum of field coverage with a 2 x2 detector configuration. With a mosaic of 4 × 2k × 2k detectors and a slightly afocal corrector (f=8500 mm, F/#=2.4) an unvignetted field of view of 43'× 43' is attained (see Fig. 2). The field is corrected in a circle of 1 degree of diameter, in order to guarantee good optical performances also at the edge of the mosaic.

The last two lenses shown in Fig. 2 are slightly separated (0.2 mm) in order to prevent them from thermal stresses in consideration of their different thermal expansion coefficients; they also act as a cryostat entrance window .

The main parameters of the two solutions are reported in Tab. 2; taking the ratio length/diameter as a figure of merit for the compactness of the instrument one can see that the second solution is the preferable.

## 2.4 Filter wheel.

The difficulty and the involved costs in obtaining filters with large format, added to the fact that smaller filters have a better homogeneity and a better optical performance, suggest the use of filter mosaic. It will be hosted inside the camera dewar, between the last lens and the mosaic itself. In such a case common commercial filters with a diameter of 60 mm can be used, with a lower cost impact. In Fig. 6 a sketch of the filter wheel aligned with the detectors is shown: the four filters match precisely the chip dimension. A supplementary hole, not drawn in Fig. 6, placed in the centre would allow the light to reach a wavefront monitoring. Another advantage of this filter wheel design is that the filters would be cooled inside the dewar and they could be correctly dimensioned as a cooled diaphragm near to the field stop (the mosaic), thus blocking part of the thermal IR contamination.

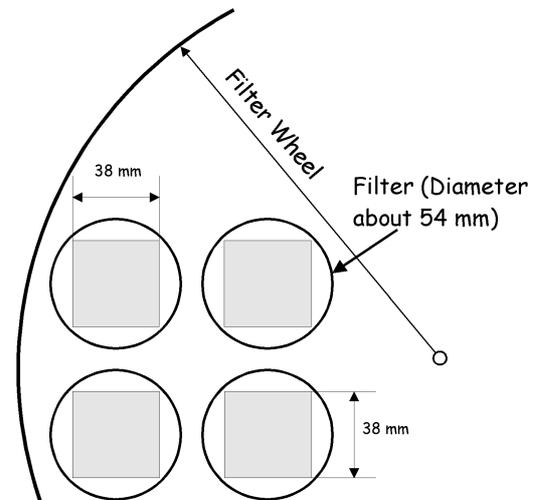

**Fig. 3:** The filter wheel matching the 4 x 2k x 2k mosaic.

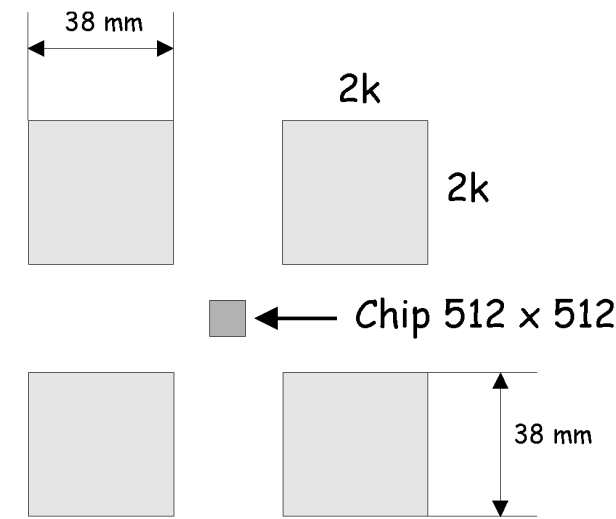
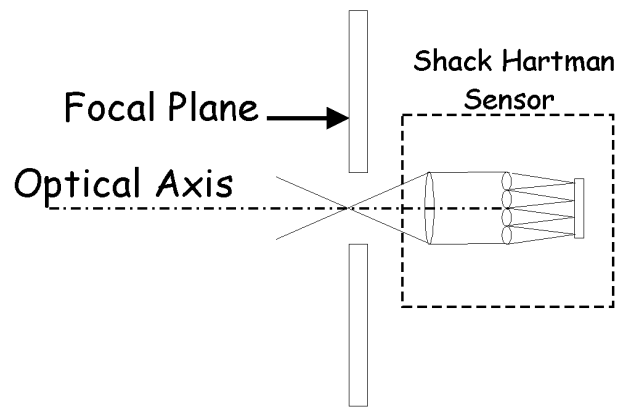

**Fig. 4:** A possible solution for the monitoring of the wavefront aberration.

## 2.5 Wavefront sensor.

The TNG telescope does not provide wavefront monitoring facility at the prime focus plane. So this has to be considered directly inside the proposed instrument.

In Fig. 4 a possible layout is drawn. It exploits the free space at the center of the mosaic. The concept is based on a Shack Hartmann wavefront sensor (SHWS) and the design offers a very simple mechanical solution and guarantees a good stability because of its symmetrical position.

The SHWS may mount a relatively small and economic chip. A common 512 x 512 pixel array working in the red wing of the visible range, where the camera has some transmissivity, should be enough for this purpose. A collimator focused on the focal plane could re-image the exit pupil on a lenslet array in order to perform the wavefront sampling.

| Mosaic | F/# | Scale | Field of view | EE(80%) | Length | First lens diameter | Back distance |
|---|---|---|---|---|---|---|---|
| 4 x 1k x 1k | 2.2 | 0.48 "/px | 24' x 24' | 10 µm | 200 mm | 136 mm | 24 mm |
| 4 x 2k x 2k | 2.4 | 0.44 "/px | 43' x 43' | 12 µm | 570 mm | 320 mm | 111 mm |

**Tab. 2**: Main characteristics of the two instrument options.

## 3. INTRUMENT PERFORMANCES

### 3.1 Limit magnitude.

In the first approximation we may calculate some performances of the instrument in terms of signal to noise ratio vs. exposure time and object magnitude. The calculation is made for a point-like source. In Fig. 5 a plot showing the SNR obtainable with a 1 h exposure time as a function of the point object magnitude is given.

The two lines represent the extreme borders of the spectral range, i. e. the dotted line for the K (2.2 µm) band and the solid one for the J (1.25 µm) band. The parameters used for the calculations are the following: trasmissivity of 0.5, sky brightness of 16 magn/arcsec$^2$ @ λ=1.25 µm and 12 magn/arcsec$^2$ @ λ=2.2 µm. The detector characteristics are those of the Rockwell Hawaii (I) HgCdTe, i. e. quantum efficiency of 0.65, read-out noise of 10 e$^-$, dark current less than 0.1 e$^-$s$^{-1}$. The band pass of the filters has been chosen of 400 nm and 290 nm for the K and J bands respectively (with unitary transmission). The exposure time is of 1 h, divided in 3600 exposures of 1 second each. All the estimations are for a seeing disk of 0.6 arcsec. It should be noticed that in both bands, and therefore in the whole spectral range from 1.2 to 2.2 µm, the instrument works, as expected, in the sky-limited region. Also the contamination flux, described in section 3.2 and 3.3, has been inserted in modelling the Fig. 5.

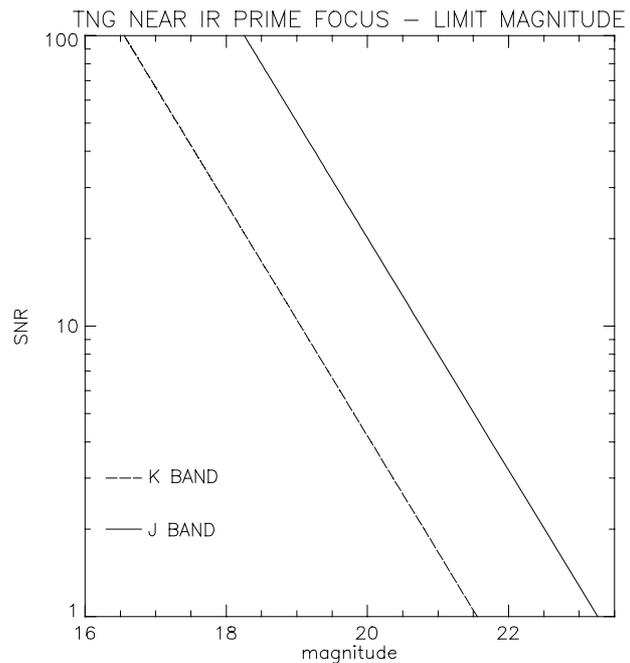

**Fig. 5:** SNR vs. object point magnitude for Texp=1 h in the extreme NIR band J e K.

## 3.2 Contamination regions.

The contamination regions are calculated for each pixel of the detector, performing a reverse ray-tracing simulation. In Fig. 6 there is a sketch of the idea. For example, the region 1 is that of the wall between the last lens and the detector, while the region 6 is the region between primary mirror and first lens. Primary mirror is the region marked 7, while the M1 central hole is the region 8.

The result of the inverse ray-tracing calculation is shown in Fig. 7. Fig. 7a is relative to a on-axis pixel, while Fig. 7b is relative to the border region of the field of view (30' from the axis). The diagrams in Fig. 7 are a projection on the plane of the semi-space seen by the given pixel. The regions of interest are those marked with 6, 7 and 8 (also visible in Fig. 6). The other regions are not relevant because they refer to cooled lenses and dewar walls. From Fig. 7 we can see how the contamination coming from region 6 is very strong. By comparison the M1 central hole gives a negligible contribute. As can be seen, the background contribution area of region 6 is comparable to that of the region 7 (the mirror). The flux coming from these areas are calculated in the next section.

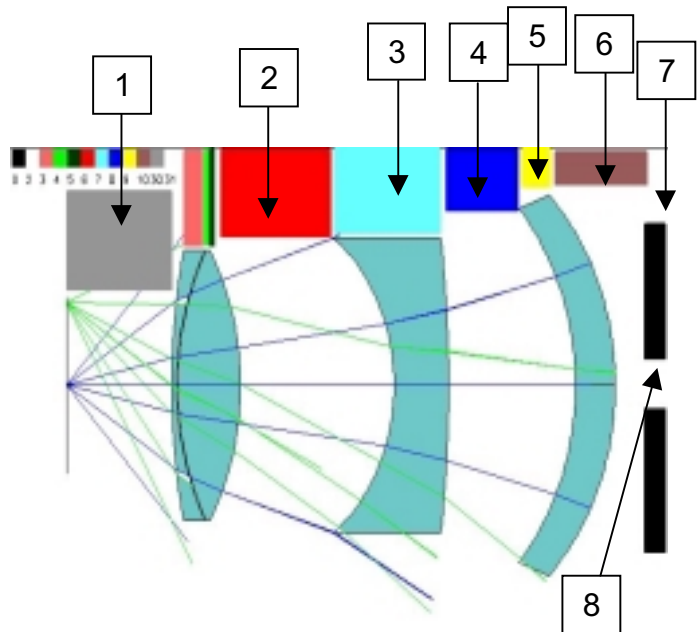

**Fig. 6:** Inverse ray-tracing calculation of the contamination flux.

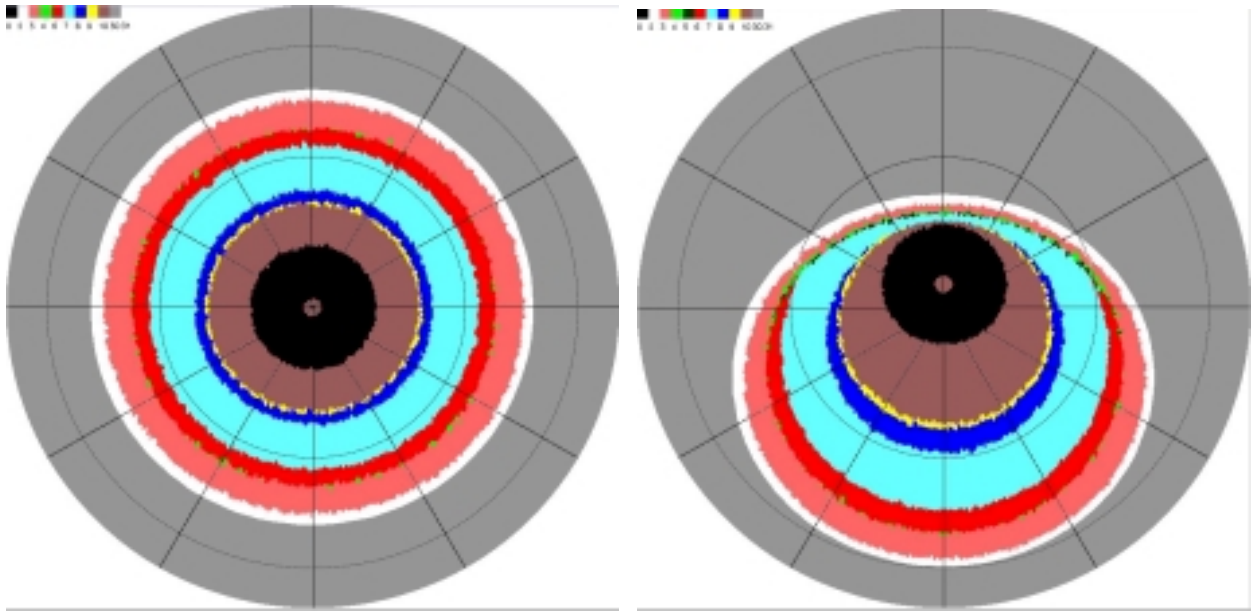

**Fig. 7:** At left: contamination for an on-axis pixel. At right that for an extreme off-axis pixel. The small central hole is the M1 obstruction, the black corona is M1, and so on for the other regions. For details on symbols see Fig. 7.

## 3.3 Background flux.

In the near infrared range the background flux plays an important role in the calculation of the limit magnitudes of the instrument. The sky brightness is greater than in the visible range and the background coming from the telescope structure becomes important approaching the K band. In fact, the intensity of the background radiation is strongly wavelength dependent. The values of the sky contribute (in magnitude/square arcsec) to the background in the NIR bands range from about 15 in J band to about 12 in K band, although these values are strongly dependent from the considered site.

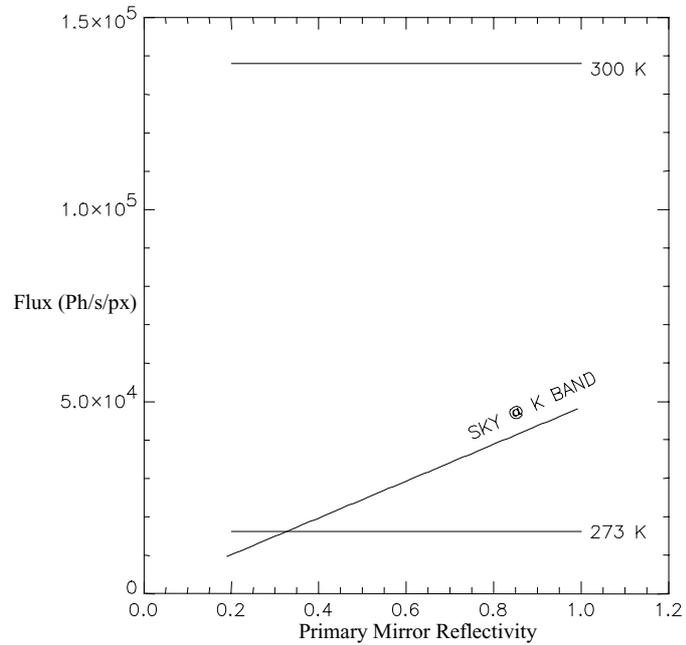

The values from the telescope background are calculated using a black body with emissivity ($\varepsilon$) 0.1 as input pupil (primary mirror). The flux from the contamination region outside the M1 has been calculated considering the region 6 in Fig. 8. The area of the region 6 is about twice that of the M1, while its black-body emissivity coefficient has been considered 1.0. The contamination flux in photon $sec^{-1}$ $px^{-1}$ due to sky and contamination regions are shown in Fig. 8. The pixel size correspond to 18×18 µm (square) with a focal length of 8500 mm, i. e. a 0.44×0.44 square arcsec. The comparison has significance only in the K band, because the sky background is dominant in the other bands. The figure shows the curve of the sky as function of the primary mirror reflectivity, which is in turn dependent from the coating status. A higher reflectivity one has just after the mirror aluminisation, while oxidation and dust present after some time are responsible for loss of reflectivity. The curve is for an 12.0 magn $arcsec^{-2}$ sky background and the flux is integrated in the whole K band (0.4 µm wide). The horizontal lines represent the fluxes from the contamination region (region 6 in Fig. 7) at the indicated temperatures, also integrated in the K band. The possible seasonal temperature excursion is evident.

**Fig. 8:** Flux (photon $sec^{-1}$ $px^{-1}$) from sky and black-body-like ($\varepsilon$=0.1) input pupil in the K band.

## 3.4 Baffling the corrector.

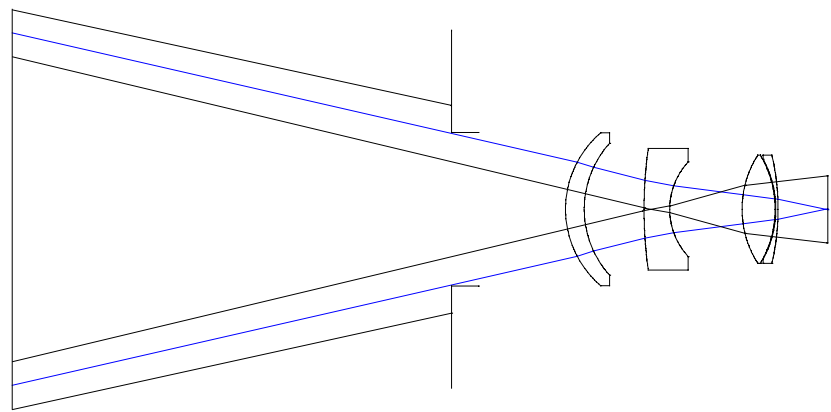

In order to better control the contamination flux visible in Fig. 8 some baffling systems are actually under study. The easiest solution seems to be obtainable by locating a diaphragm of 600 mm diameter 1 meter away from the first lens. In this way the whole region 6 is blocked, lowering consistently the contamination. The solution is depicted in Fig 9. The cost paid in this solution consists in reducing the useful mirror area to 2.5 meter diameter instead of the whole 3.5 m mirror.

**Fig. 9:** Minimum baffle system able to prevent contamination flux. Draw is not in scale.

However the vignetting factor over the field is constant, allowing a correct photometric estimation. Alternatively, one can regain mirror area increasing the diaphragm hole and reducing the distance between the chips in the mosaic. In this case a deeper but smaller field is obtained. The SNR obtained considering the solutions with and without the above described baffling system is shown in Fig. 10. With the baffle we loss about half of the mirror area (diameter from 3.5 to 2.5 m), but the contamination is dropped about 10 times. About a magnitude is gained, despite the loss of collecting area.

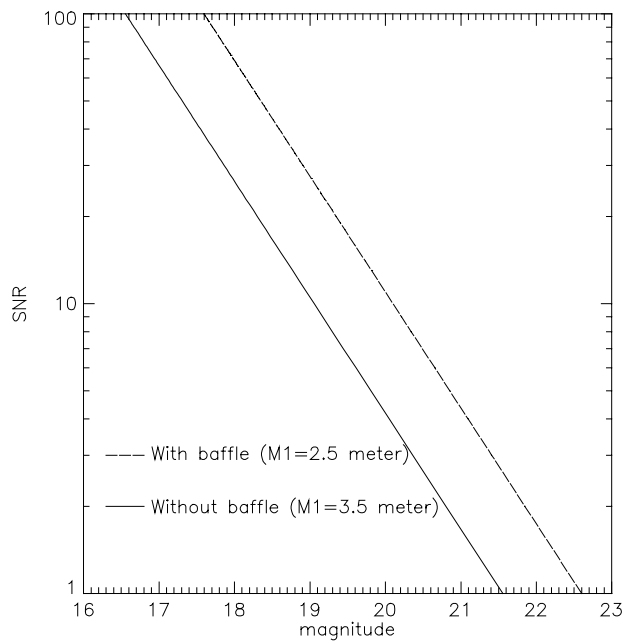

**Fig. 10:** SNR vs. magnitude comparison between the solutions with and without the baffling system depicted in Fig. 9.

## 4. CONCLUSION

A preliminary design for a near infrared ($1.2 \leq \lambda \leq 2.5$ µm) instrument for the Galileo telescope is presented. In this first study particular care has been taken to the optical design and instrument performances. We describe two options for the optical design, one exploiting a mosaic of 4 x 1k x 1k and the second for a 4 x 2k x 2k detector array. The second design seems to be more original but at the expense of larger lens curvature and aspheric coefficients. Moreover the chosen Rockwell Hawaii II 2k detectors are not yet available at the date. With the second option a field of 43' x 43' corrected for aberrations and distortions in the whole field has been obtained. In both designs a cold stop has been deliberately avoided, in order to obtain a more simple optical solution. In this way the contamination flux from regions outside the primary mirror has to be deeply studied. In fact the instrument performances has been found (as expected) to be strongly influenced by the contamination flux. A detailed calculation of it has been presented and a first baffling solution described: other solutions are presently under study.